\documentclass[sigconf]{acmart}

\usepackage{pifont}
\usepackage{tabularx}
\usepackage{color}
\usepackage{algorithm}
\usepackage{algpseudocode}
\usepackage{placeins}

\usepackage{alphabeta}  

\usepackage{hyperref}
 \hypersetup{
     colorlinks=true,
     linkcolor=blue,
     citecolor=magenta,
     filecolor=magenta,
     urlcolor=cyan,
     pdftitle={ACM DEBS 2022}
     }
\definecolor{darkgreen}{rgb}{0.1, 0.6, 0.1}
\newcommand{\cmark}{\textcolor{darkgreen}{\ding{51}}}
\newcommand{\xmark}{\textcolor{red}{\ding{55}}}

\usepackage{ifthen}
\usepackage{calc}
\usepackage{color}
\usepackage[normalem]{ulem}
\usepackage{tikz}
\usepackage{amsfonts}
\usepackage{pgfplots}

\newcommand{\annote}[3]{{
  \colorbox{#3}{\bfseries\sffamily\footnotesize\textcolor{white}{#2}}
  \color{#3}
  {\it #1}
}}

\newcommand{\commentVK}[1]{\annote{#1}{VK}{purple}}
\newcommand{\commentSV}[1]{\annote{#1}{SV}{darkgreen}}

\renewcommand{\commentVK}[1]{}
\renewcommand{\commentSV}[1]{}

\newcommand{\updated}[1]{#1}
\newcommand{\removed}[1]{}

\DeclareRobustCommand{\change}[4]{{\color{#3}#1}\ifx\relax#4\relax\else\xspace\textcolor{Gray}{\sout{#4}}\fi}

\newcommand{\FIGS}[0]{./figs}
\newcommand{\figref}[1]{Figure~\ref{#1}}
\newcommand{\secref}[1]{Section~\ref{#1}}

\setcopyright{acmcopyright}
\copyrightyear{2022}
\acmYear{2022}
\acmDOI{XXXXXXX.XXXXXXX}

\acmConference[DEBS '22]
  {16th ACM International Conference on Distributed and Event-Based Systems}
  {June 27--30, 2022}
  {Copenhagen, Denmark}
\acmPrice{15.00}
\acmISBN{978-1-4503-XXXX-X/18/06}

\title{CougaR: Fast and Eclipse-Resilient Dissemination for Blockchain Networks}

\author{Evangelos Kolyvas}
\email{ekolyvas@aueb.gr}
\affiliation{
    \institution{Department of Informatics}
    \institution{Athens University of Economics and Business}
    \city{Athens}
    \country{Greece}
}

\author{Spyros Voulgaris}
\email{voulgaris@aueb.gr}
\affiliation{
    \institution{Department of Informatics}
    \institution{Athens University of Economics and Business}
    \city{Athens}
    \country{Greece}
}
\renewcommand{\shortauthors}{E. Kolyvas and S. Voulgaris}

\copyrightyear{2022}
\acmYear{2022}
\setcopyright{acmlicensed}\acmConference[DEBS '22]{The 16th ACM International Conference on Distributed and Event-based Systems}{June 27--30, 2022}{Copenhagen, Denmark}
\acmBooktitle{The 16th ACM International Conference on Distributed and Event-based Systems (DEBS '22), June 27--30, 2022, Copenhagen, Denmark}
\acmPrice{15.00}
\acmDOI{10.1145/3524860.3539805}
\acmISBN{978-1-4503-9308-9/22/06}

\begin{document}

\begin{abstract}

Despite their development for over a decade, a key problem blockchains are still
	facing is scalability in terms of throughput, typically limited to a few
	transactions per second.
A fundamental factor limiting this metric is the propagation latency of blocks
	through the underlying peer-to-peer network, which is typically constructed
	by means of random connectivity.
Disseminating blocks fast improves not only the transaction throughput, but also
	the security of the system as it reduces the probability of forks.
In this paper we present CougaR: a simple yet efficient, eclipse-resistant,
	decentralized protocol that substantially reduces the block dissemination
	time in blockchain networks.
CougaR's key advantages stem from its link selection policy, which combines a
	network latency criterion with randomness to offer fast and reliable block
	dissemination to the entire network.
Moreover, CougaR is eclipse-resistant by design, as nodes are protected from
	having all their links directly or indirectly imposed on them by others,
	which is the typical vulnerability exploited to deploy eclipse attacks.
We rigorously evaluate CougaR by an extensive set of experiments, both against a
	wide spectrum of parameter settings, and in comparison to the current state
	of the art.

\end{abstract}

\maketitle
\section{Introduction}

Blockchains are a technology for maintaining a Byzantine fault
tolerant~\cite{byzantine} public ledger of transactions (a state machine
replication) across nodes in a Peer-to-Peer (P2P) network.
Compared to ledgers that are based on more classic consensus protocols, a key
difference is that blockchains have no central permissioning authority to
control participation of nodes in the network.
Instead, permissioning in the consensus layer is mediated by a resource.
For example, in Proof-of-Work (PoW) protocols the mediated resource is the
amount of hashing power, whereas in Proof-of-Stake (PoS) protocols it is the
amount of stake a node has that decides its participation in the network.
This mechanism of limiting each node's influence to the system by weighing its
possession of a finite resource, protects the consensus layer against sybil
attacks~\cite{sybil}.

Blockchain technology was first used as an underlying technology in
Bitcoin~\cite{bitcoin}, a cryptocurrency that was launched by Satoshi Nakamoto
in 2008.
Since then, there has been a proliferation of applications leveraging the power
of blockchains as a core component.
1st generation blockchains, like Bitcoin, introduce an electronic payment system
to transfer and store value based on cryptographic proof instead of trust.
2nd generation blockchains, like Ethereum~\cite{ethereum}, provide a
Turing-complete programming language that can be used to encode arbitrary state
transition functions simply by writing a few lines of code in a smart contract.
3rd generation blockchains, like Cardano~\cite{cardano}, improve upon the
previous two generations by solving three big pain points:
scalability, interoperability, and sustainability.

Despite their fame and evolution, a key problem blockchain systems are still
facing today is scalability.
Although transactions per second (tps) is not the most accurate measure, as the
size of the transactions may vary drastically (transaction bytes per second may
be a better measure), it can provide an overview to compare the most popular
representatives of each blockchain generation with a classical payment system:
Bitcoin can support a maximum of 7 tps, Ethereum 15 tps, and Cardano 7 tps,
which are in stark contrast to established payment systems like Visa that can
support more than 2000 tps~\cite{tps}.

The low throughput of these systems constitutes the focus of many proposals
aiming at improving it,
including sharding~\cite{elastico, omniledger, rapidchain, monoxide},
alternative consensus mechanisms~\cite{bitcoinNG, algorand, byzCoin, ouroboros,
snow},
changes in the way and structure of how data is stored~\cite{segWit,
bitcoinCash, cmpctBlock},
using a directed acyclic graph (DAG) instead of a chain~\cite{spectre, phantom,
dagcoin, iota, byteball, nano},
or employing payment channels~\cite{lightningNet, raidenNet, hydra},
side chains~\cite{plasma, peggedSidechains, liquidityNet},
and cross-chain protocols~\cite{cosmos, polkadot}.
While all these proposals offer sophisticated solutions, a fundamental factor
limiting the performance of blockchain systems is the latency of the layer
underneath, that is, the inherent message propagation delay introduced by the
P2P network.

Reducing the message propagation delay can lead to higher transaction
throughput, as it allows one to increase the block size, to increase the block
generation rate, or to employ faster consensus algorithms.
Besides higher throughput, reducing the propagation delay also strengthens the
security of the system by lowering the probability of \emph{forks}.
A fork is the situation where two blocks happen to be generated in parallel
(i.e., neither of the two miners being aware of the other block while generating
their own), leading to a temporary ambiguity on what the official state of the
chain is.
As such ambiguities may be exploited for illicit behavior, minimizing message
propagation delay does not only offer higher performance but also stronger
security guarantees.

At its core, a blockchain protocol functions by periodically combining
transactions into \emph{blocks} and broadcasting them over the network.
Block dissemination implementations are typically based on unstructured overlay
networks, formed based on random connectivity:
each node establishes a number of connections to a random set of peers.
A typical example of such a network is the Bitcoin
network~\cite{propagationInBitcoin}.
However, it is easy to see why such a policy is suboptimal:
a protocol that does not take neighbors' proximity (in terms of network delay)
into account may result into delivering a block to a node within the same
datacenter through a path that spans the entire planet.

As a consequence, proposals have been made for faster and more sophisticated
dissemination protocols~\cite{kadcast, perigee}.
While all these solutions provide some speed improvements, they turn to handle
the issue as a trade-off between fast and secure (eclipse-resistant)
dissemination of blocks.
According to them, the dissemination should be:
\emph{a)} either fast (but not secure), by employing a scoring function that
turns to match the well-connected peers among themselves~\cite{perigee}.
However, such an adaptive protocol that can be manipulated by an
adversary~\cite{tendrilstaller} and leave the victim just with 1 or 2
non-adversarial neighbors, thus eclipsing the vast majority of its connections,
\emph{b)} or secure (but not fast), by employing a performance-agnostic
protocol~\cite{propagationInBitcoin, kadcast} which almost completely disregards
any tuning to be faster.

Another issue with blockchain networks is bandwidth consumption.
A well-designed dissemination protocol should be bandwidth efficient for block
relay, otherwise it can fail to achieve its goals.
In a dissemination protocol which nodes carelessly waste their bandwidth by
relaying much redundant information (e.g. flooding the network each time they
meet a new block), many things can go wrong.
In a bandwidth inefficient protocol, downstream peers can have moderate inbound
bandwidth spikes, however upstream peers may have significant outbound
bandwidth spikes, especially the nodes that receive the new block in the early
stages of dissemination, before their neighbors~\cite{cmpctBlock}.
Upon receiving a block earlier than its neighbors, a node needs to send the new
block multiple times, one to each neighbor.
Such bandwidth spikes not only delay the transmission of blocks, but also can
make consumer-grade internet connections temporarily unusable.
Thus, decreasing bandwidth consumption in blockchain networks is an important
factor to enhance scalability as it is beneficial for many individuals running
nodes, which, in turn, enhances security indirectly.

In this paper we focus on building more efficient P2P topologies for block
dissemination tailored for state-of-the-art blockchains.
We present CougaR:
a simple yet efficient adaptive decentralized protocol that decides which
neighbors a node should connect to based on a combination of proximity (in terms
of network latency) and randomness.
CougaR is not only fast, but also eclipse-resistant and bandwidth-efficient.
We advocate our proposed protocol by presenting an extensive simulation-based
evaluation demonstrating its performance.

The remainder of this paper is organized as follows.
We first present a short background on epidemic dissemination in \secref{sec:background}.
In \secref{sec:design} we advocate our design and we present the CougaR protocol.
In \secref{sec:experimental_setup} we lay out the experimental setup and in
\secref{sec:evaluation} we present an extensive evaluation of
CougaR with respect to its performance for a wide range of parameter settings.
In \secref{sec:comparison} we present related work and we experimentally compare
our protocol against a number of state-of-the-art dissemination protocols for
blockchain systems.
Finally, \secref{sec:conclusions} concludes the paper.

\section{Epidemic Dissemination Background}
\label{sec:background}

Epidemic protocols for data dissemination have been extensively studied in the
past, leading to the identification of \emph{push} and \emph{pull} as the two
main representatives.
Both methods assume that every node may initiate communication to peers selected
\emph{uniformly at random} out of all other participating nodes.

\subsection{Push-based Dissemination}

In \emph{push-based dissemination}, when a node receives a message it has not
seen before, it instantly forwards it to a number of other nodes, which in turn
do the same.
Due to the reactive nature of this operation, new messages spread exponentially
fast to a significant portion of the network.

The push paradigm is very efficient in the early stages of dissemination, when
most nodes are still unaware of the new message, thus forwarding it to
arbitrarily chosen nodes is likely to spread it further.
However, it suffers in later stages of dissemination, when most nodes have
already received this message, therefore forwarding it arbitrarily will most
likely deliver it to an already informed node, wasting network resources for no
gain.

Even worse, as nodes have no control over \emph{who} should forward the new
message to them, some nodes may never receive a given message simply because no
other node chose to forward it to them.
To alleviate this shortcoming, push-based dissemination schemes often employ
high levels of redundancy, so that the probability of any one node being left
out diminishes, at the cost of very high network overhead.
Kermarrec et. al~\cite{kermarrec} report that each node should forward a message
to around 15 other nodes to probabilistically achieve complete dissemination,
using network resources that are in the order of 15-fold higher than the
theoretical optimal of delivering a message to each node once.

\subsection{Pull-based Dissemination}

In \emph{pull-based dissemination}, nodes periodically contact arbitrary other
nodes to ask whether a new message is available, and to pull it from them if so.
Due to its proactive nature and periodic polling, pull-based dissemination does
not spread messages as fast as its push-based counterpart, most notably in the
early stages of dissemination when most polls do not bring any news.
However, as each node is responsible for fetching new messages to itself,
eventually every single node receives the message, i.e., no node is ``left
out''.
Moreover, periodic polling messages aside, the pull-based strategy is very
network efficient, as every new message is delivered exactly \emph{once} to each
node.
Its moderate dissemination speed, though, renders it inapt for use as-is in
blockchain systems.

\begin{table}
  \centering
  \begin{tabular}{|m{0.24\linewidth}|m{0.1\linewidth}|m{0.1\linewidth}|m{0.39\linewidth}|}
  \hline
                               & \textsc{Push}  & \textsc{Pull}  & \textsc{Key Mechanism}              \\ \hline
  \textsc{Dissemination Speed} & \cmark\ fast   & \xmark\ slow   & Reactively forwarding messages upon reception \\ \hline
  \textsc{Reliability}         & \xmark\ no     & \cmark\ yes    & Having control over \emph{who} sends you messages         \\ \hline
  \textsc{Network Overhead}    & \xmark\ high   & \cmark\ low    & Delivering messages once per node \\ \hline
  \end{tabular}
  \caption{Push vs. Pull}
  \label{tab:pushpull}
\end{table}

Table~\ref{tab:pushpull} summarizes the pros and cons of push-based and
pull-based epidemic dissemination, along with the key mechanisms inducing each
property.
What we need is a dissemination model combining the advantages of both worlds.
We describe this model in the following section.

\section{Protocol Design}
\label{sec:design}

Designing a data dissemination protocol involves two parts.
First, providing a \textbf{link placement strategy}, that is, deciding which
links should be established between nodes to be used for dissemination.
Second, devising the \textbf{dissemination model}, that is, defining the
specific interactions between nodes that allow messages to be efficiently and
reliably disseminated over the available links.

In the following sections we define our proposed dissemination model, followed
by our proposed link placement strategy and the detailed protocol operation.

\subsection{Dissemination Model}
\label{sec:dissemination_model}

We model our dissemination network as an undirected graph $G(V,E)$, where $V$ is
the set of vertices, or \emph{nodes}, and $E$ is the set of undirected edges, or
\emph{links}, among nodes.
Two nodes are called \emph{neighbors} when there is a link connecting them.
Links are bidirectional, and each node can arbitrarily select a number of other
nodes (known as its \emph{outgoing neighbors}) to establish links to.
Two neighbors are equally responsible for forwarding new blocks to each other,
irrespectively of who took the initiative to establish the link between them.
Thus, blocks are being disseminated by being forwarded across links in either
direction.

When a node forwards a block to one of its neighbors, we refer to the sending
node as the \emph{upstream peer} and to the receiving one as the
\emph{downstream peer}.
In the context of another block, their roles may be reversed, should the block
traverse their link in the opposite direction.

Our dissemination model borrows from both the push-based and the pull-based
models to achieve the best of both worlds.
It adopts reactive message forwarding from the push-based model to cater for
fast dissemination, and policies from the pull-based model to guarantee
reliability and to keep network overhead low (Table~\ref{tab:pushpull}).

More specifically, with respect to reactive message forwarding, when a node
validates a new block it immediately advertises it to all its neighbors.
This behavior, attributed to the push-based model, satisfies the first mechanism
of Table~\ref{tab:pushpull} and constitutes the key ingredient for fast,
reactive dissemination.
In contrast, links are established in a proactive manner, asynchronously with
respect to the dissemination of blocks, as discussed in
\secref{sec:link_placement}.

Link bidirectionality helps alleviate a well-known reliability issue associated
with selecting links exclusively in a single direction.
If nodes select only their downstream peers (e.g., as in the push model), a node
may be left without any upstream peers, failing to receive blocks.
Likewise, if nodes only select their upstream peers, some nodes may be left
without downstream peers, failing to disseminate blocks they produce.
By establishing that every single node is entitled to set up a number of links
to nodes of its choice, and that these links are used for propagating blocks in
\emph{both directions}, no single node is left without downstream or upstream
dissemination paths.
This satisfies the second key mechanism of Table~\ref{tab:pushpull},
guaranteeing that for any arbitrary node communication can flow in both
directions: \emph{from it} and \emph{to it}.

\begin{figure}[t]
  \includegraphics[width=\columnwidth]{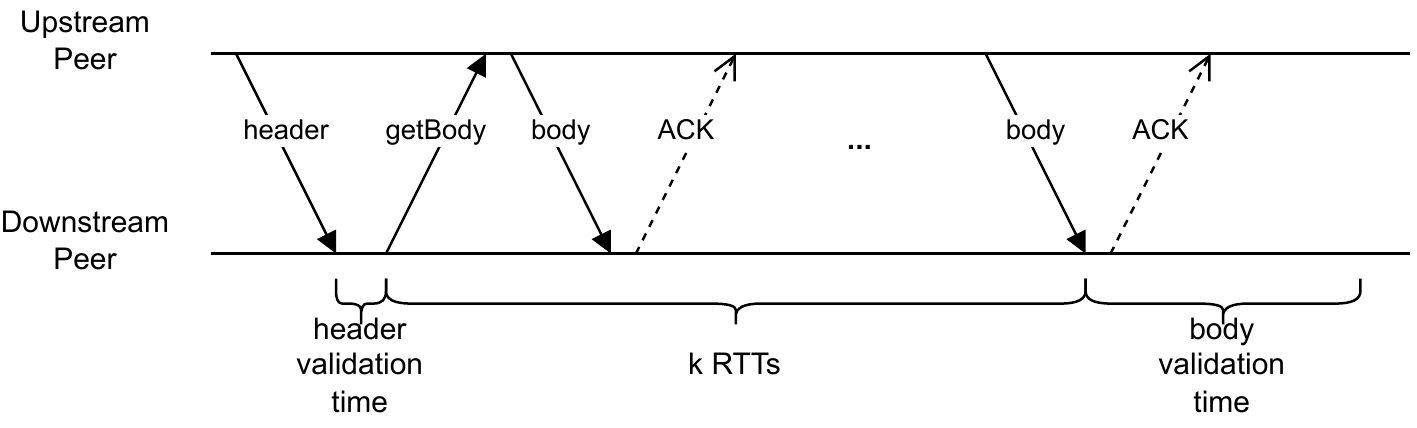}
  \caption{CougaR's forwarding scheme}
  \label{fig:headerBody}
\end{figure}

Finally, with respect to keeping network overhead low
(Table~\ref{tab:pushpull}'s third mechanism), we adopt a two-step block
forwarding scheme that gives the receiving node control on how many copies of a
block it wants to receive.
A node acquiring a new block forwards only a \emph{digest} of the block (rather
than the block itself) to all its neighbors, letting each of them individually
decide whether it would also like to be sent the actual block, on demand.
It is, thus, the receiver's decision to pull the actual block from one or more
of the neighbors that provided the corresponding digest.

Although this is a generic mechanism for keeping network overhead low, the
specific anatomy of blocks can be leveraged to further optimize the whole
process.
Blocks consist of two parts, a \emph{header} and a \emph{body}.
The header is small\footnote{The header is 80B in Bitcoin and below 1KB in
Ethereum and Cardano} and can be safely assumed to fit within a single IP
packet.
The body is generally orders of magnitude larger\footnote{The body is up to 2MB
in Bitcoin and in the order of 30KB in Ethereum and Cardano}.
In our model, the header itself serves as the block digest.
First, the sending node already has the header; there is no need for any extra
digest to be produced on demand.
Second, and most importantly, the receiving node can perform a validity check on
the header before requesting the actual block body, diminishing the effect of
DoS attacks that try to spread invalid blocks.
Upon receiving the actual block body, a node performs a complete validity check
for all transactions of the block, and provided the check succeeds the node is
ready to start forwarding it further to its own neighbors (modulo the one it
received it from) in the same way.
\figref{fig:headerBody} illustrates CougaR's forwarding scheme, requiring $k$
round trips for the body of the propagated block to be received.

\subsection{Link Placement Strategy}
\label{sec:link_placement}

\begin{figure}[t]
  \includegraphics[width=\columnwidth]{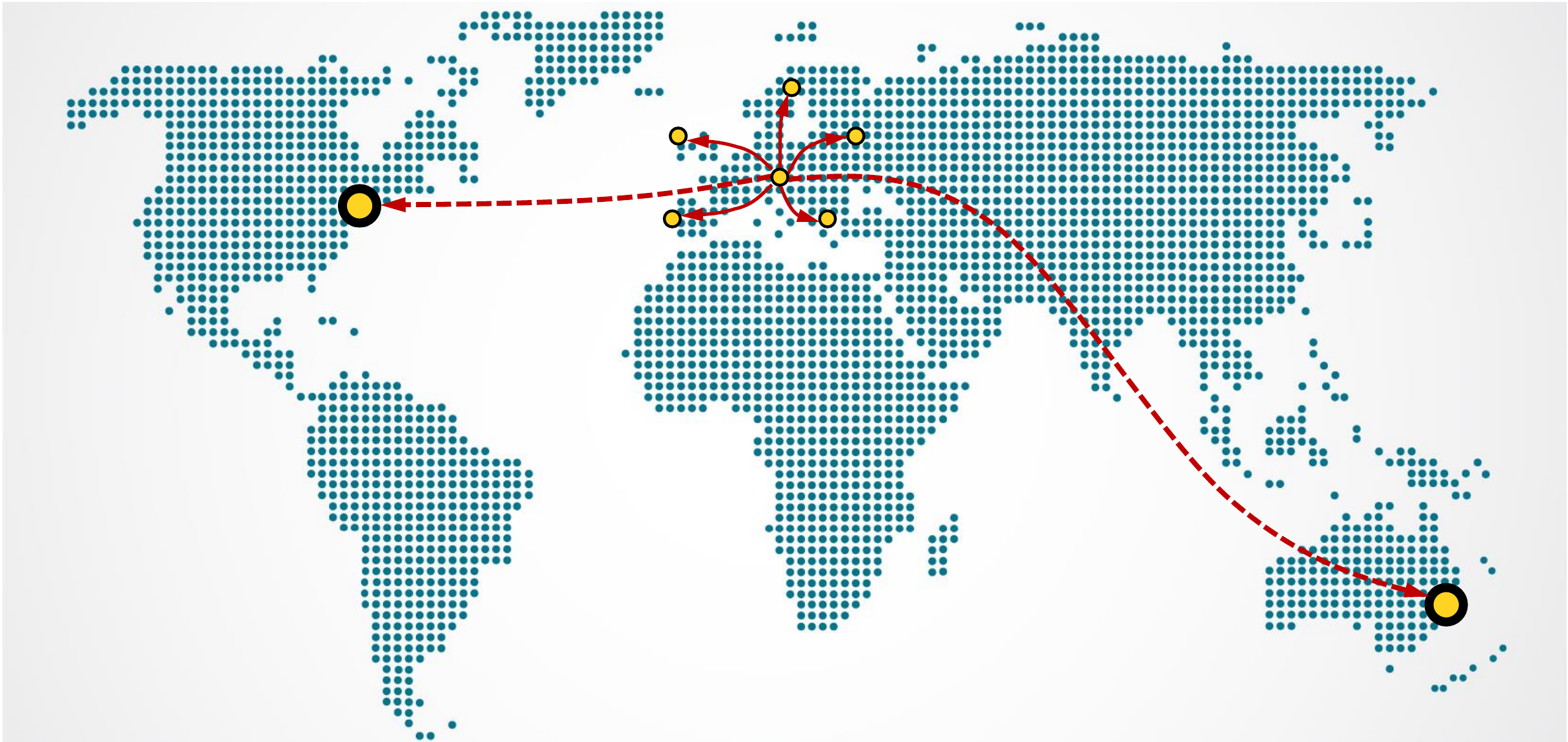}
  \caption{Dissemination overview: Blocks should be propagated to a few nearby and to a few distant nodes.}
  \label{fig:world}
\end{figure}

Fast dissemination at a global scale is a two-faceted endeavour.
First, a new block should be disseminated fast and exhaustively at a local
scope, harnessing the low-latency links of geographically proximal locations and
ensuring that every single nearby node receives the block through a fast,
low-latency, local path.
Second, a dissemination algorithm should also encompass a global outlook,
managing to spread the news fast to distant locations.

Intuitively, a block generated in Amsterdam should reach a node located in
Zurich fast, over local links, rather than over a long-distance path that first
visits Sydney.
At the same time, the node in Sydney should also receive the block relatively
fast, over a path that contains a direct shortcut from somewhere around
Amsterdam to somewhere in Australia, rather than waiting for the block to slowly
cross all of Europe and Asia over many local dissemination steps.

\figref{fig:world} illustrates an overview of the proposed strategy.
Blocks should be forwarded both across local, low-latency links, and distant,
long-range ones.
This brings up a fundamental question: how to pick short-range and long-range
links.

Kleinberg~\cite{kleinberg} defines a small-world model that facilitates
efficient routing over short paths in a large network of nodes without global
knowledge.
Although routing and dissemination are different problems, we could leverage the
observations on routing to build overlays for efficient dissemination.
However, Kleinberg's model assumes nodes organized in a regular grid structure,
which does not reflect the actual Internet topology.
Additionally, this model relies on the reliable measurement of nodes' distances,
and on establishing links mostly to close nodes, but also to fewer nodes of
increasingly higher distances.
Explicitly selecting \emph{distant nodes}, though, involves two risks.
First, by deterministically opting for highest-latency nodes as neighbors, nodes
located in isolated areas (e.g., remote islands in the middle of an ocean), will
result into a highly biased topology and link distribution, with a very high
number of links to them.
Second, and even worse, as being a ``distant'' node can easily be emulated by
merely delaying all communication, malicious nodes could easily take advantage
of this to attract links from the entire network, something we want to prevent
at all costs.
In contrast, the property of being ``close'' to a given node cannot be faked.

In our proposed protocol, we form links based on metrics that cannot be faked.
Namely, each node establishes links to a number of other nodes out of the
following two sets:
\begin{itemize}
\item \textbf{Close neighbors:} These are nodes exhibiting low network latency
	to each other. Intuitively, such nodes tend to be geographically close to
		each other, although our protocol is location agnostic and is concerned
		exclusively with network latency.
\item \textbf{Random neighbors:} These are neighbors picked uniformly at random
	out of all participating nodes. The rationale behind this decision is that
		our protocol does not demand a multitude of nodes at each distant
		region, but rather just a few sporadic representatives. Therefore, if
		every node forwards a block across a few \emph{random} links, the block
		should quickly get widely dispersed across the world, albeit at a sparse
		density. Forwarding, subsequently, to close neighbors bridges the gap,
		turning a block's sparse distribution into a dense, exhaustive
		dissemination reaching every single participating node across the globe.
Last but not least, placing random links creates overlays resembling random
		graphs, which are known for their low diameter and extreme resilience to
		failures.
\end{itemize}

Turning the dissemination model and link placement strategy detailed above into
a protocol able to operate in a global-scale distributed environment is a
non-trivial task.
A number of issues should be taken into account, most notably the crucial
adaptivity and self-healing features to let it operate flawlessly in dynamic
conditions and arbitrary failures inherent in real-world settings.
Such conditions include joining and leaving nodes, fluctuating network
performance, and dynamic node load.

\subsection{The CougaR Protocol}
\label{sec:protocol_operation}

We propose an adaptive decentralized protocol, which works as follows.
Each node establishes $C+R$ links to nodes of its choice and measures their
latencies by a number of ping messages.
Periodically, a node discards its $R$ most distant neighbors, replacing them by
$R$ randomly picked ones.
In case multiple neighbors are close to the node, not sufficiently separated in
the latency space, it picks one of them at random.
Each node $v$ is also free to impose a locally determined \emph{degree limit} of
$M_v$ (with $M_v>C+R$) links in total (including the $C+R$ links established by
itself), reflecting $v$'s bandwidth and ability to handle a number of
connections in parallel.
That is, if a node $u$ attempts to establish a link to a node $v$ whose degree
limit $M_v$ has been reached, $v$ refuses and $u$ picks another node at random.
Algorithm~\ref{alg:placement} shows the pseudocode of our link placement
algorithm.

\begin{algorithm}
\caption{Link Placement Algorithm}
\label{alg:placement}

\begin{algorithmic}[1]

\State \textit{// PSS = the underlying peer sampling service}
\State \textit{// responsive = is both alive and able to handle another connection}
\State
\State \textit{// Rejuvenate node's outgoing neighbor set (ONS)}
\Loop{} periodically
	\ForAll{node $v \in ONS$}
		\State measure the RTT to node $v$
	\EndFor
	\State
    \While{$ONS.size > C$}
        \State remove the node with the highest RTT from the $ONS$
    \EndWhile
    \State
    \While{$ONS.size < C+R$}
        \State pick a random node $v$ from $PSS$
        \If {$v$ is responsive}
            \State add $v$ to the $ONS$
        \EndIf
    \EndWhile
\EndLoop

\end{algorithmic}
\end{algorithm}

The network overlay emerging from this simple decentralized protocol possesses a
number of desirable properties.
First, as each node establishes $R$ bidirectional links to random other nodes
(not counting the $C$ links to its closest neighbors, or links established by
other nodes to oneself), the resulting overlay has far more edges than the
respective family of \emph{$R$-regular random graphs}, which are known to be
\emph{a.a.s. connected}\footnote{``asymptotically almost surely'' connected} for
$R\ge3$~\cite{randomGraphs}.
Therefore, the resulting overlay is infinitely scalable with respect to
connectedness.
Second, as each node has a lower degree bound of $C+R$ bidirectional links, no
node can be isolated, as these links alone account for $C+R$ downstream and
upstream dissemination paths.
Third, each node's degree also has an explicit upper bound, which prevents the
(accidental or intentional) scenario of a node ending up with too many
connections, rendering it unable to serve them all in an efficient manner.
Last, but not least, the periodic rejuvenation of a node's neighbor set helps it
adapt to dynamic conditions, replacing non-responsive or distant nodes by
randomly picked ones, and to maintain links to nearby neighbors of low network
latency.
In effect, at any given time a node maintains links to $C$ nodes of the
\emph{close neighbors} set and $R$ nodes of the \emph{random neighbors} set, as
defined in \secref{sec:link_placement}.

Assuming that latency proximity cannot be faked by the attacker, and that the
underlying peer sampling service~\cite{peerSampling} can provide nodes with a
truly \emph{unbiased} random set of peers, CougaR is an eclipse-resistant
protocol as attackers have no means of arbitrarily manipulating the set of
established connections.
That is, neither of the ``close'' and ``random'' properties can be faked.
We also assume the presence of an effective third-party defense mechanism
against sybil attacks, so that a physical node cannot acquire an unlimited
number of identities.

Block propagation follows the dissemination model detailed in
\secref{sec:dissemination_model}.
A node acquiring a new block forwards its header to all its neighbors.
Upon receiving a header from a neighbor, a node validates it locally and
subsequently requests the corresponding body from that neighbor.
Should more neighbors advertise the same header before the node has and
validated the body in question, the node may spawn additional body requests in
parallel.

CougaR introduces parameter $P$, controlling the degree of parallelism by
setting the maximum number of body requests that may be pending on behalf of a
node for a given block at any given moment.
Setting $P$ to its lowest value, 1, results into the most bandwidth-sensitive
setup, known as the \emph{conservative policy}, where a node requests a body
only from a single neighbor, the one that delivered the respective header first.
On the other end, setting $P$ equal to a node's number of neighbors leads to the
\emph{greedy policy}, in which a node requests the block from \emph{all} nodes
that sent it the respective header, until a download completes and the received
block has been validated.
Intermediate values of $P$ offer the flexibility to fine-tune the trade-off
between bandwidth conservation on the one side, and faster download speed with
higher redundancy on the other.

\section{Experimental Setup}
\label{sec:experimental_setup}

We split our evaluation up into two parts.
\secref{sec:evaluation} evaluates our protocol in a wide range of parameter
settings.
\secref{sec:comparison} compares CougaR against the state-of-the-art, while
discussing its novelty and differences in comparison to related work.
The experimental setup presented in the current section applies to both parts.

All evaluation was performed in the Peer-Net Simulator~\cite{peernet}, a
discrete-event simulator for P2P protocols written in Java, as a fork of the
popular PeerSim simulator~\cite{peersim}, able to execute protocols not only in
simulation mode but also in real networks.

In all evaluation, the notation C$x$-R$y$ denotes a configuration with $x$ close
and $y$ random outgoing links established by each node.

\subsection{Topologies and Network Latencies}
\label{sec:wondernetwork}

Obtaining realistic measurements dictates the use of realistic data.
Network latencies play a central role to the accuracy of our protocols'
assessment.
Therefore, we used the following method to compile a dependable real-world
latency trace.

First, we acquired the latency trace made publicly available by
WonderNetwork~\cite{wondernetwork}.
This trace reports the round-trip times across all pairs of around 250 servers
distributed across 87 countries in all continents, measured repeatedly for over
two weeks, in November 2021.
A worth-mentioning detail about this latency trace is that it is asymmetric.
That is, the time it takes for a message to be sent from node A to node B is not
necessarily equal to the time it takes for the message to be sent from node B to
node A.

Second, we collected the geographic locations of nodes for
Bitcoin\footnote{\url{https://bitnodes.io/}},
Ethereum\footnote{\url{https://ethernodes.org/}},
and Cardano\footnote{\url{https://adapools.org/}}.

Then, we mapped each node to its closest WonderNetwork server by estimating
distances based on polar coordinates.
As expected, more than one nodes could be mapped on a single WonderNetwork
server, corresponding to nodes operating in the same city (or possibly
datacenter).
This resulted into three distinct datasets, differing in how many times they
included each WonderNetwork server, reflecting the geographic distribution of
nodes in the respective blockchain.

Finally, we projected the three aforementioned datasets to three new datasets of
16,000 nodes each, maintaining a proportional node distribution.
These latency datasets were used to run experiments in our evaluation.
However, due to space limitations, we only present Bitcoin topology results.
Ethereum and Cardano node topologies present very similar results, and have, thus, been omitted.

\begin{figure*}
  \includegraphics{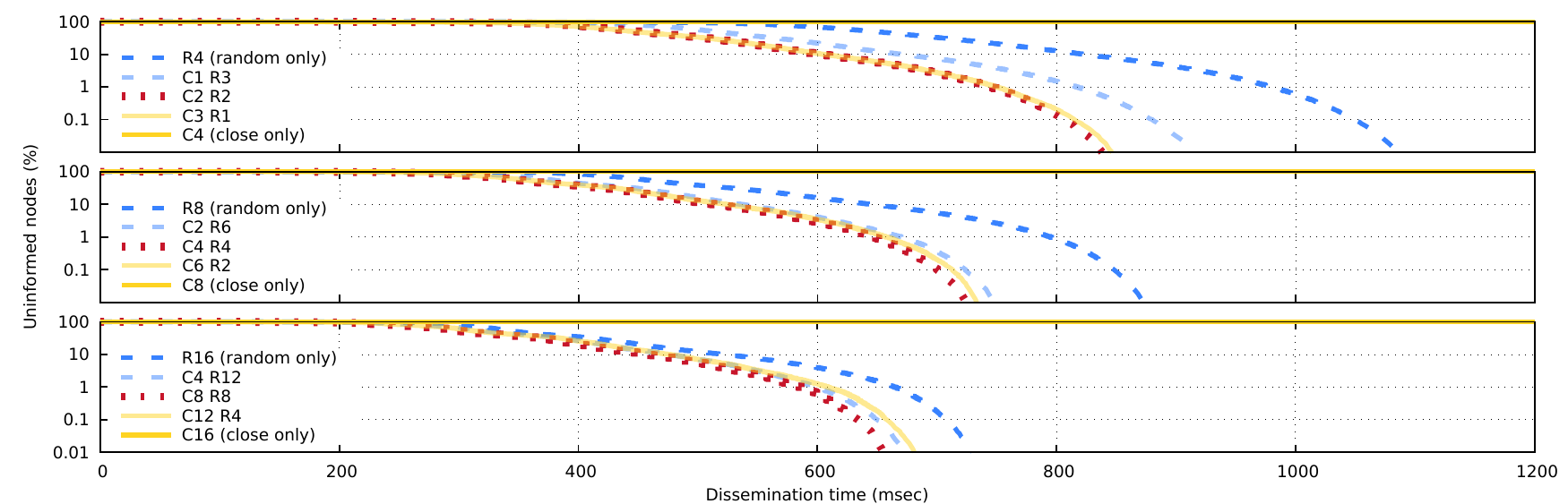}
  \caption{Selecting \emph{all} random links (\textbf{R}, dark blue dashes) or
	\emph{all} close links (\textbf{C}, dark yellow lines) yields the worst
	performance irrespectively of the node degree. \emph{Mixing} random and
	close links (faded shades) gradually speeds up dissemination. Adopting
	\emph{equal shares} of both (red dots) always provides the fastest (or
	practically as good as the fastest) dissemination.}
  \label{fig:multiplot}
\end{figure*}

\subsection{TCP Considerations}
\label{sec:tcp}

We assume that nodes communicate over the TCP protocol.
In order to acquire more accurate estimates of block transfer times, we take a quick look at TCP's operation, notably on its congestion avoidance mechanism.

TCP is a reliable, connection-oriented communication protocol that allows bidirectional communication between two nodes.
When two nodes establish a TCP connection, they set (among other things) an initial window size for congestion avoidance in each direction.
This typically corresponds to a small multiple of the MSS (Maximum Segment Size, i.e., the maximum number of bytes TCP can fit in a single IP packet, with a default value of 536).
This window size dictates up to how many bytes can be sent by the sender before receiving an acknowledgement from the receiver.
That is, when the sender needs to send data to the receiver, it sends a small number of packets back-to-back, in one go.
Upon receiving an acknowledgement, the sender infers that there was no congestion along the path to the receiver, so it increases its window size typically by some fixed amount of bytes (additive increase).
Should an acknowledgement get lost or delayed, the sender assumes there is congestion, so it decreases the window size to a fraction of its current value (multiplicative decrease).

In terms of the time it takes to transfer $L$ bytes from a sender to a receiver, if $L$ is smaller than the initial window size all $L$ bytes will be sent in a single batch of packets, taking around RTT/2.
Else, part of the $L$ bytes will be sent on the first batch, to which the receiver will respond by an acknowledgement (ACK), which will trigger the sender to transmit the second batch of packets.
The time for the ACK to travel back to the sender and the next batch to propagate to the receiver is yet another RTT.
In general, a data chunk requiring an extra $k$ batches to be transferred on top of the initial batch, will take $\frac{1}{2}+k$ RTTs to complete \updated{(see \figref{fig:headerBody}).

TCP connections are established by a three-way handshake, asynchronously to block dissemination, and are kept alive for as long as the respective nodes remain neighbors.
As different TCP implementations use different initial window sizes and increment/decrement steps, our experiments are concerned with how many extra RTTs are needed for transferring a block between two nodes, rather than with the exact number of bytes transferred in each batch.
}

\section{Standalone Evaluation}
\label{sec:evaluation}

We evaluate CougaR by investigating its performance across a wide range of
parameter settings, including alternative options of the link selection policy,
different node degrees, a range of block validation delays and block sizes, and
different levels of parallelism, as presented in the following sections.

\subsection{Link Selection: Close vs. Random}

The first parameter to examine was the effect of the link selection policy, by
means of the ratio of close/random links established by nodes.
We ran a number of experiments to evaluate all combinations of close and random
links for different node (outgoing) degrees.
In these and all following experiments, unless otherwise stated, we fixed the
header and body validation delays to 5\,msec and 50\,msec per node,
respectively, as reported for
Bitcoin\footnote{\url{https://statoshi.info/d/000000003/function-timings}}.

\figref{fig:multiplot} shows the progress of dissemination in the course of time
elapsed since a block's generation, by indicating for each point in time the
percentage of nodes that have not yet received and validated the respective
block.
The figure contains three plots, corresponding to 4, 8, and 16 links per node,
respectively.
Each line corresponds to a distinct experiment and shows the average of the
dissemination of 100 blocks originating at uniformly randomly chosen miners
for the Bitcoin node topology.
Dark blue dashes represent \emph{random only} setups, while dark yellow lines
represent \emph{close only} setups.
Gradually fading colors indicate the gradual mixing with links of the
alternative type, while red dots correspond to the equal sharing between
close and random links.

We observe that, for all degrees checked, splitting the links evenly between
close and random ones tends to give either the fastest, or negligibly off the
fastest dissemination speed.
Exclusive use of random or close links, on the other hand, yields the worst
performance in all cases.
We also observe that, in any node degree, the setups involving only close nodes
fail to reach all nodes, as the network becomes disjoint into disconnected
components due to the nodes' greedy policy to team up exclusively with nearby
nodes.

\subsection{Link Selection: Average Node Degree}
\label{sec:node_degree}

\begin{figure}[t]
  \includegraphics{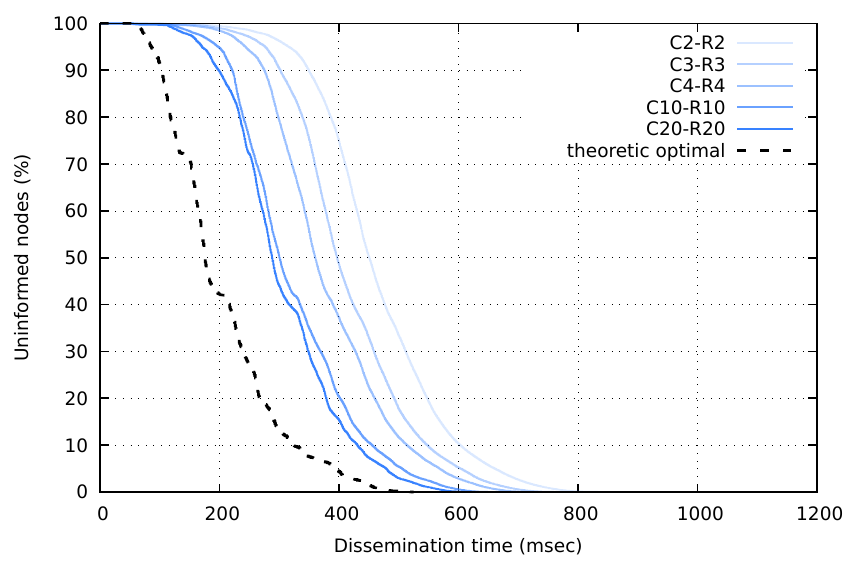}
  \caption{Dissemination time vs Node degree}
  \label{fig:degree}
\end{figure}

The next parameter to investigate was the effect of the average node degree,
that is the number of links each node is entitled to establish with its peers.
We know that in Bitcoin every peer establishes 8 outgoing connections to a
random set of peers, while peers set an upper limit of 117 incoming connections
each~\cite{bitcoinTopology}.
In Ethereum, each peer maintains up to 25 connections, up to 13 of them being
outgoing connections\footnote{We consider geth, the most prevalent Ethereum
client.}.
In Cardano, each node establishes 5 to 20 outgoing connections\footnote{We
consider Cardano Shelley implementation.}.

\figref{fig:degree} shows the dissemination progress in time for a number of
different node degrees in the Bitcoin topology.
There is a trade-off while tuning the number of links per node:
A higher number of outgoing connections per node results into faster dissemination
speed, however we need more bandwidth to maintain these connections.
Additionally, we observe that the benefits of further increasing the node degree
become negligible after some point.
For example, switching from four connections (C2-R2) to six connections (C3-R3) has
a greater impact in dissemination speed than switching from 20 connections
(C10-R10) to 40 (C20-R20).

In addition, \figref{fig:degree} plots a \emph{theoretic optimal}.
This theoretic optimal corresponds to an imaginary scenario in which every node
has unlimited resources and is able to forward every block to all other nodes in
a single hop.
By comparing the performance of the 40-connections scenario (C20-R20) to that of the
theoretic optimal (16K connections per node and unlimited bandwidth) we see that
they lie within the same order of magnitude.
Indeed, the former reaches 95\% of the network only 1.44 times slower than the latter.
This constitutes a strong indication that there is not much to gain by pushing node
degree beyond a certain level.

Consequently, we consider Bitcoin's choice of letting nodes select 8
connections each a reasonable one, and we fix this value for the rest of the paper.

\subsection{Effect of Block Validation Delay}

\begin{figure}[t]
  \includegraphics{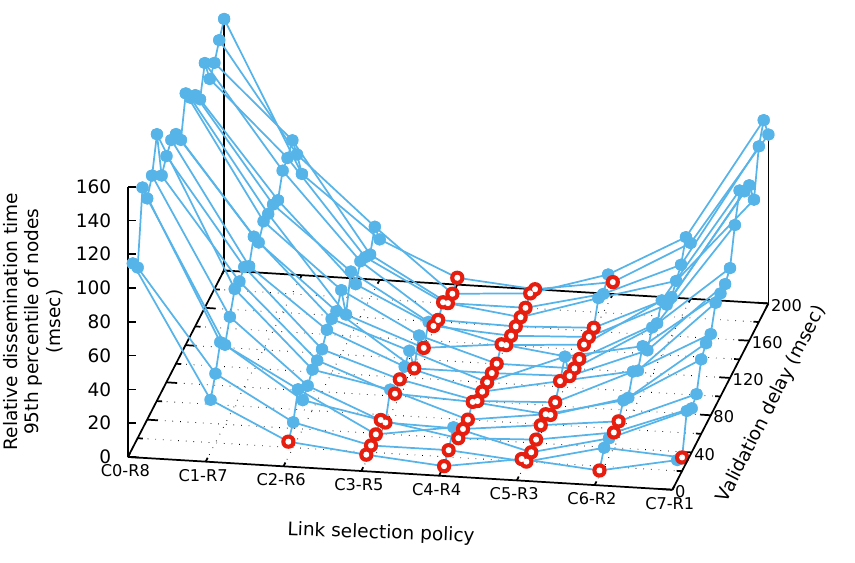}
  \caption{Relative dissemination time among all link selection policies and
    validation delays on the 95th percentile of nodes. Red circles indicate the
	link selection policies that are within 10\,msec off the optimal one, per
	validation delay.}
  \label{fig:cr_proc}
\end{figure}

As explained in \secref{sec:dissemination_model}, a node acquiring a block does
not push it further until it has locally validated it.
This is done to prevent the propagation of malformed blocks in the network.
The validation process involves a series of checks both on the header and on the body of
each block, and takes non-negligible time to perform.

Header validation checks whether the header satisfies the respective Proof-of-Work or
Proof-of-Stake eligibility criteria,
whether the header links to a valid past header on the chain,
whether the timestamp is valid, and so on.
Typically these constitute a fixed number of checks and require a short amount of time.
We fixed the validation delay for headers to 5\,msec for all our experiments as
the processing time of the header is usually independent of the size and the
processing time of the body.

Body validation, on the other hand, depends
on the number of transactions in a given block and on their respective complexity.
A node has to check each individual transaction of a block by confirming 
its syntactical correctness and by verifying its execution validity.
Transactions involving smart contract calls can prove
far more CPU intensive compared to those that simply transfer assets between
wallets.
The aggregated effect of sequential validations across multi-hop dissemination
paths can significantly increase the total dissemination time.

This brings up the following question:
Does the per-node validation delay have an effect on which link selection
policy provides the optimal results?

We ran experiments for an extended set of validation delays and close/random ratio
combinations, for node degrees ranging from 6 to 20, all of which indicate that
the equal splitting of links between close ones and random ones always yields the optimal
(or negligibly close to the optimal) performance.

\figref{fig:cr_proc} shows one representative of these sets of experiments,
namely the one for 8 links per node and body validation delay ranging from 10 to
200\,msec for the Bitcoin node topology.
We measured, for each validation delay, the time each combination of close and
random links took to disseminate blocks to 95\% of the nodes, on average.
Presented times are relative to the dissemination time of the fastest link
selection policy across the same validation delay.
That is, for each validation delay we identified the fastest dissemination
time, and we subtracted it from the dissemination times of all link selection
policies, to highlight their relative performance.

Red circles highlight those link selection policies that perform best or are
within 10\,msec off the best, per validation delay.
It is clear that the policy pertaining to the equal sharing of close and random
links can be trusted as an optimal pick independently of the validation
delay.

Note that the link selection policy of close links only (C8-R0) is not shown in
this plot, as when nodes focus exlusively on their nearby neighbors the emerging
overlay is seggregated into a number of disjoint components, obstructing
dissemination altogether.

\subsection{Effect of Block Size}

\begin{figure}[t!]
  \includegraphics{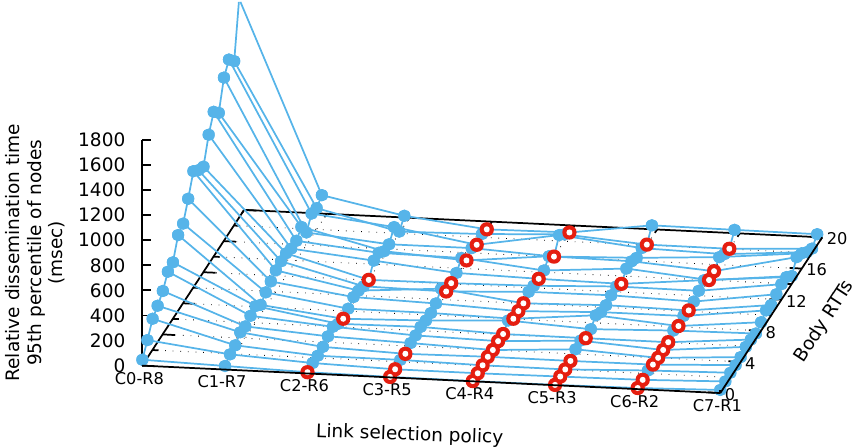}
  \caption{Relative dissemination time among all link selection policies for
	each number of body RTT transfers on the 95th percentile of nodes. Red circles
	indicate the link selection policies that are within 10\,msec off the optimal
	one, for each number of RTTs.}
  \label{fig:cr_bRTTs}
\end{figure}

In this section we focus on different block size scenarios, and we investigate
how they affect the optimal link selection policy.
We start by assessing their effect on the time required to transfer a block from
one node to another.

Let \emph{RTT} be the round-trip time between two nodes, $A$ and $B$, and let
$A$ have a block it wants to forward to $B$.
Per our dissemination model (\secref{sec:dissemination_model}) this transfer
will take place in two steps.
First $A$ will forward the header to $B$, and if $B$ has not already received
the block body from another source it will send back to $A$ a request to pull
the body, which $A$ will subsequently send to $B$.
This will account for a total time of \emph{at least} 1.5 \emph{RTT}, assuming
the block body is small enough to fit in the initial TCP window between $A$ and
$B$.
As outlined in \secref{sec:tcp}, TCP will split up the body transfer into $k$
(with $k\ge 1$) discrete batches of packets, depending on the body size and the
TCP window size adaptation policies.

To assess the effect of the extra RTTs incurred by larger blocks on
dissemination, we ran experiments ranging the body RTTs required per transfer
from 1 to 19.
For each RTT value, we ran a number of experiments for all possible combinations
of close and random links, for node degrees ranging from 6 to 20.
In all these experiments we observed that adopting an equal share of close and
random links gives either the best or negligibly off the best results, thus
confirming that our proposed link selection policy is a good choice.

\figref{fig:cr_bRTTs} (similar in style to \figref{fig:cr_proc}) shows a
representative set of these experiments, namely the one for 8 links per node.
The figure reports the results of experiments for each combination of link
selection policy and RTTs-per-transfer, when dissemination reaches 95\% of the nodes.
Presented times are relative to the dissemination time of the fastest link
selection policy across a given RTTs-per-transfer value.

Red circles highlight those combinations in the parameter space that are either
optimal (for a given RTTs-per-transfer value) or are no more than 10\,msec off
the optimal performance.

For the rest of the evaluation we consider only setups where the links each node
is allowed to establish are equally split between latency-wise close peers and
random peers.

\subsection{Bandwidth Efficiency}
\label{sec:headerBody}

\begin{figure}[t]
  \includegraphics{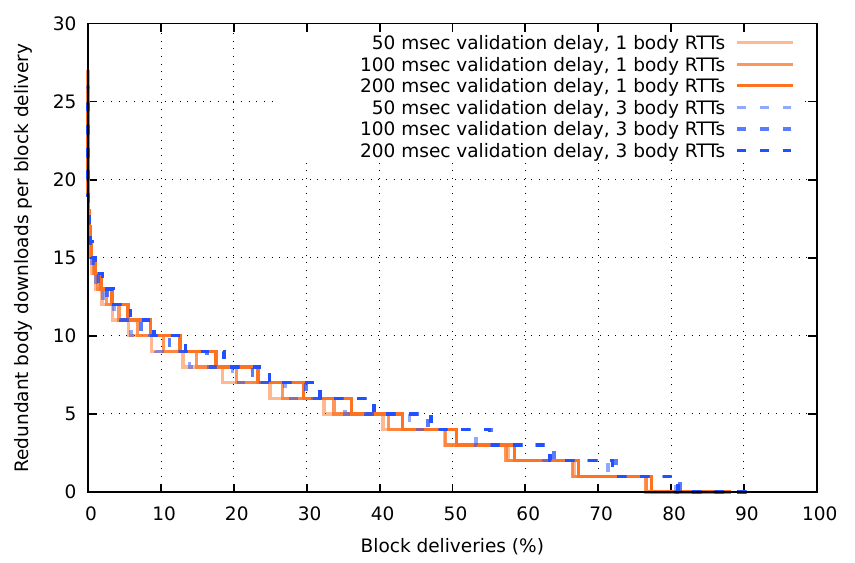}
  \caption{Greedy approach: Redundant body downloads per block delivery, for all deliveries of a total of 100 blocks to all 16K nodes.}
  \label{fig:BW_usage}
\end{figure}

Many blockchain protocols implement various methods to keep bandwidth utilization within reasonable levels.
Preventing the waste of network resources is not only important for the network itself, but also for reducing the unnecessary load on blockchain nodes and letting them act and communicate more rapidly when needed.

Our protocol adopts a block relay scheme in which blocks are forwarded in two steps:
the header is pushed first; the body is pulled then, on demand.
It is, thus, the receiving node that is in control of which and how many of its neighbors to pull a body from in parallel,
pulling from the first node only (the conservative approach) to pulling from all until a download completes and the received block has been validated (the greedy approach).

\begin{figure}[t]
  \includegraphics{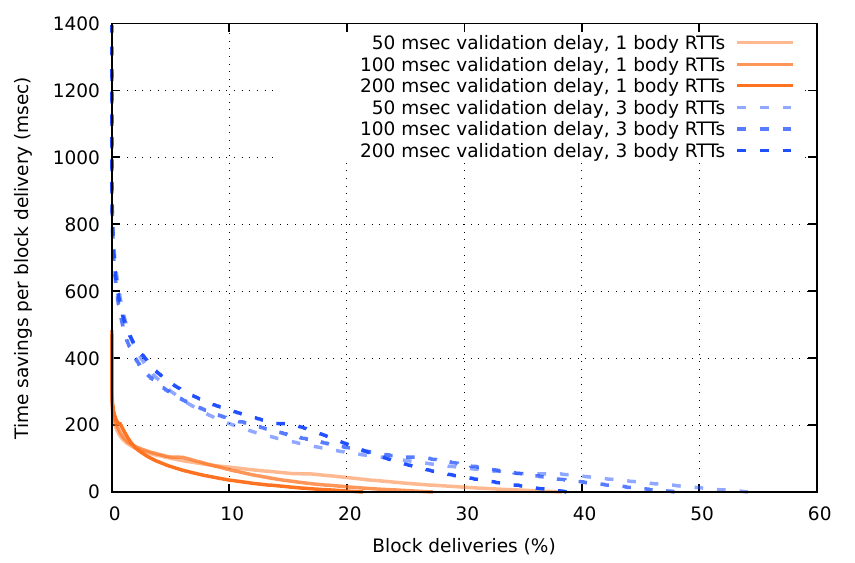}
  \caption{Greedy vs Conservative: Absolute body transfer time gains per block delivery, for all deliveries of a total of 100 blocks to all 16K nodes.}
  \label{fig:time_savings}
\end{figure}

In order to assess the extra bandwidth consumed by the greedy approach, we
carried out a number of experiments both using the greedy and the
conservative approaches.
Each experiment involved 1,600,000 block deliveries: 100 blocks, each being
delivered to all 16,000 nodes.
We recorded the number of times a body was pulled by each node, on average.
\figref{fig:BW_usage} reports the number of redundant (i.e., more than one) body pulls
per block delivery, with block deliveries sorted in a descending redundancy order.

Three validation delay values (50, 100, and 200\,msec) and two block sizes
(needing 1 and 3 RTTs to be transferred) were considered, resulting into six
scenarios.
We notice a substantial degree of redundancy in all six scenarios, with larger
block sizes resulting into slightly more redundant pulls.
This makes sense, as larger blocks take longer to be delivered, in which period
the receiver gets to send pull requests to more of its neighbors.

In the worst case, all 6 combinations tend to request over 15 extra times the
block body, which is in fact the average node degree, as each node connects to 8
peers, so (on average) another 8 nodes connect to it.

\figref{fig:time_savings} presents the respective time savings each individual
of the aforementioned 1.6M block deliveries observed with the greedy approach in
comparison to the conservative one.
Scenarios exhibiting large blocks gain more performance benefits from the greedy
approach.

\begin{figure}[t]
  \includegraphics{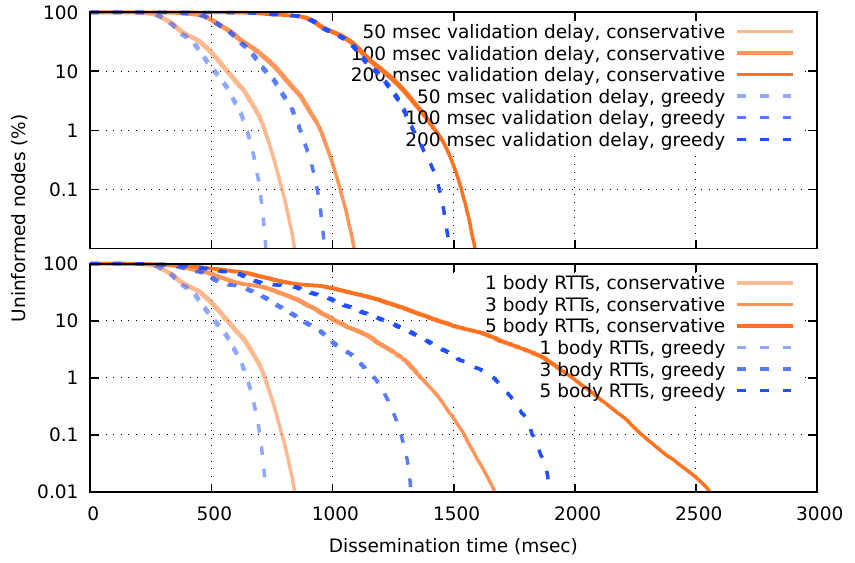}
    \caption{Greedy vs Conservative: Dissemination evolution over time}
  \label{fig:conservative_greedy}
\end{figure}

\figref{fig:conservative_greedy} presents the evolution of dissemination by
presenting the percentage of nodes remaining uninformed in the course of time for
the greedy and the conservative approaches.
The benefit gained by the greedy approach appears to be constant irrespectively of the
per-node validation delay (top).
However, the number of extra RTTs required for larger block sizes has a
clear correlation to the performance gains earned by the greedy approach (bottom).

\begin{figure}[t!]
  \includegraphics{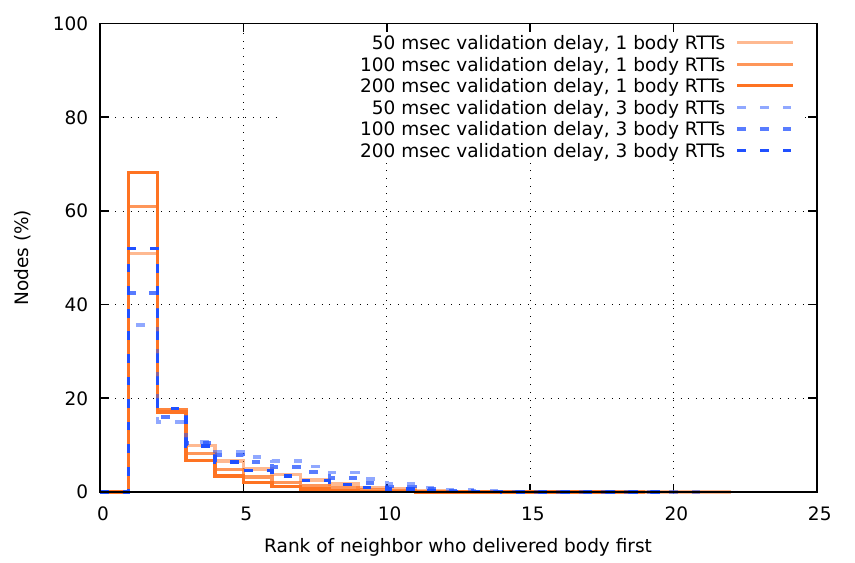}
  \caption{Greedy approach: Which peer manages to send the body first}
  \label{fig:golden_header}
  \vspace{-10pt}
\end{figure}

To explore the trade-off between high dissemination performance and low
bandwidth usage, we investigated the ideal number of body download requests a
node should spawn in parallel.
More specifically, we executed the greedy version of the protocol (i.e., with an
unlimited number of download requests until a download completes and the
received block has been validated),
and we recorded which neighbor was the one that succeeded in delivering the body
\emph{first}, in terms of the order it was asked.
I.e., whether it was the neighbor asked first, second, third, and so on. 
The respective distribution is presented in \figref{fig:golden_header}.

In \figref{fig:golden_header} we observe that an overwhelming percentage of
downloads are served successfully by the first four upstream peers asked.
Therefore, we choose to fix CougaR's parallelism parameter $P$ to 4 for the
comparison to related work presented below.

\section{Evaluation against Related Work}
\label{sec:comparison}

Having completed the standalone evaluation of CougaR, we now proceed to comparing it against state-of-the-art baseline algorithms.

\subsection{Related Work}
\label{sec:baselines}

A number of techniques have been employed in real systems or proposed by the research community for block dissemination in blockchain systems, presented in the following sections.

\subsubsection{\textbf{Random}}
\label{sec:baseline_random}

The random connection algorithm is the simplest and most widely deployed
connection policy in blockchains.
The most notable example using this algorithm is Bitcoin.
When bootstrapping, a node is not aware of any other peers of the network.
It will use a DNS seeder to reach at least one of them.
The node, subsequently, gossips with the peer(s) it already knows to learn
addresses of additional peers and to advertise its own address.
As mentioned in \secref{sec:node_degree}, in Bitcoin each
node establishes 8 outgoing connections to randomly picked other nodes and
accepts up to 117 incoming connections set up by other nodes.
Just like CougaR, the random connection algorithm is eclipse-resistant,
provided a peer sampling service that discovers an unbiased random set of peers.

While randomly formed topologies are known for their low diameter and resilience
to failures, they suffer from suboptimal path delays.
This is due to the fact that they do not take other nodes' latency-wise proximity
into consideration.
Thus, in large networks nodes will choose, with high probability, distant neighbors,
significantly growing path delays.

\subsubsection{\textbf{Geographic}}
\label{sec:baseline_geographic}

A simple heuristic, proposed in Perigee~\cite{perigee}, to improve the random connection algorithm is to take geographic
locations of nodes into account, assuming these can be inferred through their IP addresses.
Such a heuristic would pick some geographically close neighbors and some distant ones.

However, this algorithm has practical difficulties.
First, since many nodes (over half in Bitcoin~\cite{bitnodes}) connect through
a Tor network~\cite{tor}, IP addresses cannot be known.
Second, with the help of VPNs or proxies, one may present an IP address
in an arbitrary geographic region.
As such a link selection algorithm is prone to manipulation,
it cannot be considered eclipse resistant.

In order to compare CougaR against this algorithm, we
implemented a protocol that establishes links based on nodes' geographic locations.
We split the nodes up into six sets based on the continents they are located in:
Africa, Europe, North America, South America, Oceania, and Asia.
Each node picks half of its neighbors randomly among peers from the same
continent and the other half randomly among peers from different continents.

\subsubsection{\textbf{Structured Overlay}}
\label{sec:baseline_kademlia}

A recent work~\cite{kadcast} proposes a broadcast protocol based on
the Kademlia~\cite{kademlia} DHT.
Harnessing DHT properties, it achieves dissemination in a
logarithmic number of hops.
However, its performance is only slightly better compared to systems based on
random topologies of equal node degree, which does not justify the extra
overhead of building and maintaining Kademlia.

In \secref{sec:comp_eval} we compare such an algorithm with CougaR.
Note, however, that CougaR adopts a constant node degree independently of the
network size, whereas in Kademlia node degrees grow logarithmically
with the size of the network.
We consider such a structured topology as eclipse resistant under two
conditions:
\emph{(a)} each node has an unforgeable DHT identifier, and
\emph{(b)} the overlay has a mechanism to locate the alive node whose ID is
closest to a desired point in the ID space.

Ethereum operates based on the Kademlia DHT too.
In Ethereum, however, Kademlia is only used as a membership management
protocol, i.e., to discover other nodes and to pick neighbors, while block
dissemination is performed over an unstructured P2P overlay in an epidemic
fashion similar to that of Bitcoin.
Nodes establish up to 13 outgoing connections and accept up to 12 incoming connections each.
These connections are used to disseminate both transactions and new blocks.

\subsubsection{\textbf{Score-based}}
\label{sec:baseline_perigee}

Perigee~\cite{perigee} proposes a scoring function to assess every neighbor
based on its ability to deliver blocks, and retains the ``best'' subset of
neighbors at regular intervals.
Each node also periodically connects to random new peers to explore potentially
better-connected neighbors.
Each node maintains 8 outgoing connections and accepts up to 20 incoming
connections.

A pitfall of such a protocol is its defense against eclipse
attacks~\cite{eclipseBitcoin, eclipseEthereum}, as an adversary could easily
dominate a victim's connections by providing well-connected peers.
A well-known way of performing this attack is by an
attacker skipping the block validation process in order to be the first to deliver new
blocks to the victim, and acquire a top score in its ranking.
It only takes a handful of colluding nodes to dominate the scoring links of the victim.
Consequently, Perigee and similar protocols cannot be considered eclipse resistant.
Moreover, as we see in \secref{sec:comp_eval}, it is still questionable how
such a protocol can handle a moderate percentage of block transfer failures that
can confuse the scoring mechanism, making it pick, in essence, peers at
random.

\subsubsection{\textbf{Blockchain Distribution Networks}}
Another line of work proposes high-speed \emph{Blockchain Distribution Networks}
(BDNs)~\cite{bloxroute, falcon, fibre} to help nodes propagate blocks and
transactions faster.
These solutions, however, are not fully decentralized and rely on a trusted
relay network.
A malicious actor can potentially attack such networks by performing a
person-in-the-middle attack.
Besides that, if a blockchain system accepts the trust assumptions made by such
proposals, our proposed protocol could leverage the proximity (in terms of
network latency) properties these systems provide to achieve even faster
dissemination.

\subsection{Comparison to Related Work}
\label{sec:comp_eval}

\begin{figure}[t]
  \includegraphics{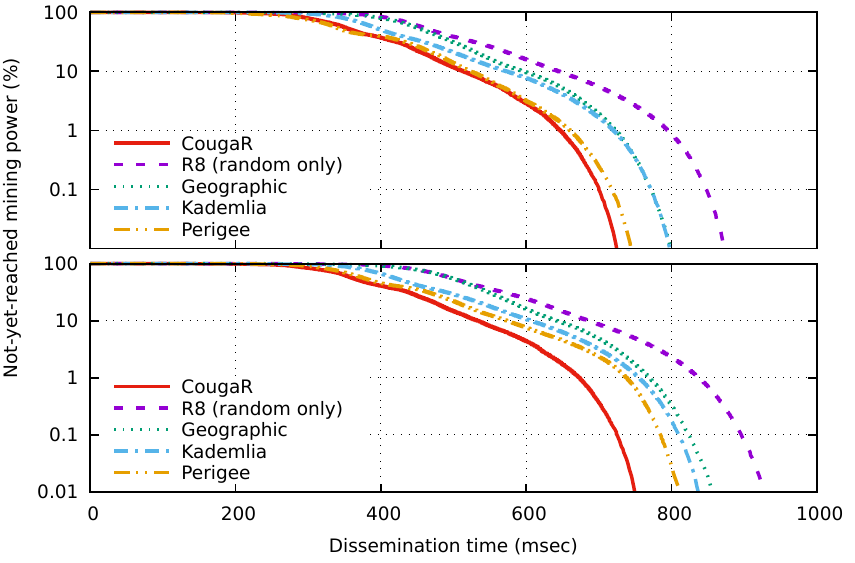}
    \caption{Protocol comparison with uniform mining power. No failures (top), 10\% block transfer failures (bottom).}
  \label{fig:comparison_unif}
  \vspace{-10pt}
\end{figure}

\begin{figure}[t]
  \includegraphics{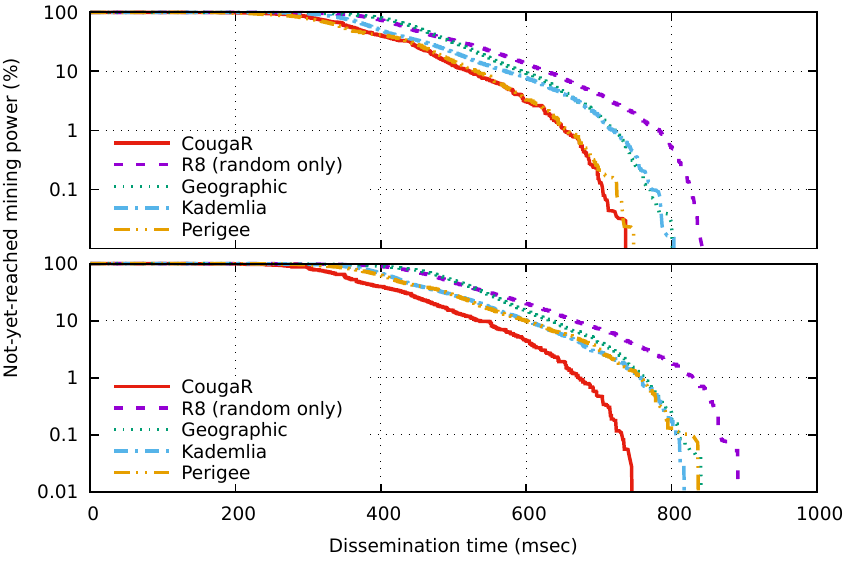}
  \vspace{-20pt}
  \caption{Protocol comparison with exponentially distributed mining power. No failures (top), 10\% block transfer failures (bottom).}
  \label{fig:comparison_expn}
  \vspace{-14pt}
\end{figure}

We implement C4-R4 and we set the parallelism parameter $P=4$, limitting the number of concurrent body pull requests to four:
two from our close-set and two from our random-set, starting by requesting the body
from the first two peers that delivered to us the respective header from each set.
To prevent malicious behavior~\cite{delayAttack}, we set a timeout on each body pull request equal to
twice the number of RTTs required to transfer the body (\figref{fig:headerBody}).
When this timeout expires for a pull request, we send a new request to another peer
having advertised the respective header.
Block delivery is considered complete when we have received and fully validated
the body.
From the remaining of this section we will assume CougaR to be configured with this policy.

We compare CougaR with:
\emph{(a)} the random connection algorithm (R8), in which every node connects with
eight random nodes (\secref{sec:baseline_random}),
\emph{(b)} the geographic connection algorithm, in which every node
connects to four random nodes from the same continent and four random ones from
other continents (\secref{sec:baseline_geographic}),
\emph{(c)} the structured overlay algorithm, based on Kademlia
(\secref{sec:baseline_kademlia}), and
\emph{(d)} Perigee (\secref{sec:baseline_perigee}), which implements a scoring
function that lets nodes calibrate their neighbor selection every 100 blocks by
replacing the ``worst'' pair of peers (\emph{subset} of size two, in Perigee terminology)
by two random nodes.
As Perigee is an adaptive protocol, we ran the protocol for 12.8K blocks
(128 rounds of 100 blocks each) to let it converge before measuring our metrics, as suggested in that paper.
For each algorithm we collect the results taken from the dissemination of 100 blocks,
generated by random nodes.

\figref{fig:comparison_unif} considers a scenario with default values for all parameters.
That is, 5 msec header validation delay, 50 msec body validation delay, 1.5 RTTs for
a block to be transferred (header $\rightarrow$ pull-request $\rightarrow$ body),
and a uniform mining power distribution (i.e., each node is equally likely to generate a block).
The upper part plots the results when no transfer failures occur,
while the lower part plots the results when 10\% of block transfers fail,
to assess the protocols' behavior in the face of failures.

\figref{fig:comparison_expn} considers the same scenario, except for nodes'
mining power (be it \emph{hashing power} for PoW or \emph{stake} for PoS)
now following an exponential distribution.
The upper part reports the case of no transfer failures,
while the lower part reports the results when 10\% of block transfers fail.
Block dissemination (vertical axis) is reported in terms of the percentage
of mining power reached.
Obviously, in the case of uniform mining power (\figref{fig:comparison_unif}),
this is equivalent to the percentage of nodes reached.

Both uniform and exponential mining power distribution give clear rankings
of algorithm performance.
As expected, Random connectivity performs the worst in all cases.
Kademlia and Geographic share the third and fourth positions, with Kademlia
having an edge when failures occur due to its structured topology.
Perigee comes closer to CougaR, mostly in the absence of failures.
When block transfers exhibit random failures, CougaR clearly outperforms
Perigee.
This is further emphasized for an exponential mining power distribution,
where Perigee performs equally to Kademlia.
This is due to Perigee's scoring functions being sensitive to transfer
failures, in the presence of which it proposes sub-optimal nodes as neighbors.
In case of an exponential mining power distribution, this phenomenon is
even more prominent as the scoring function fails to locate the relatively
few peers generating most of the blocks.

An interesting fact that is clearly illustrated in \figref{fig:comparison_unif}
and \figref{fig:comparison_expn} and deserves to be highlighted, is the difference
between CougaR and Geographic.
One could (wrongly) assume that picking random long-distance and random
short-distance links (i.e., the Geographic heuristic) should be practically equivalent
to CougaR's selection of close and random links.
This is strongly disproven by the performance comparison.
The difference lies in the fact that it is not \emph{geographic distance}, but
\emph{latency-wise proximity} that CougaR takes into account for link placement.
Likewise, picking random peers is not equivalent to picking explicitly distant peers either.
As an example of the inefficiency of the Geographic protocol, we can think of a
node located in Lisbon picking a neighbor in Tromsø as a node from the same
continent, while discarding a node in Rabat as a node from a different continent.

Bandwidth consumption is yet another dimension that should be highlighted.
CougaR does not only outperform Perigee in terms of dissemination speed,
but also in terms of bandwidth consumption.
CougaR needs almost a constant small number of body deliveries per node to
achieve this performance, independently of the node degrees. 
Perigee, on the other hand, requires nodes to request and download the body of every single
block from literally \emph{all} their neighbors who provided the respective
header, carrying on even after a body has been received and validated,
in order to properly rank them by means of the scoring mechanism.

Another difference between CougaR and Perigee lies in the convergence time
to reach their peak performance.
CougaR converges asynchronously with respect to block dissemination,
as latency measurements between nodes are independent of block propagation.
On the other hand, Perigee needs many rounds, of hundreds of blocks each, to converge
and reach a reasonable performance.
Thus, CougaR provides a calibrated overlay almost from the beginning, not after
many thousands of blocks have been generated.

Last but not least, a very important advantage CougaR has to offer over Perigee has to do with the
security and the protection against being eclipsed.
In Perigee, an attacker can launch an eclipse attack by providing blocks
earlier than other nodes to the victim, thus dominating its set of neighbors.
The only mechanism, in Perigee, mitigating this attack is the selection of
two random neighbors, which, in case of eight outgoing connections per node,
constitutes the 25\% of its links.
The rest 75\% of the links can be easily eclipsed~\cite{tendrilstaller}.
In contrast, CougaR is by design shielded against this attack vector,
as nodes select their close neighbors prioritizing on low network latency,
a property that cannot be forged if the attacker is not for real in
network proximity to the victim.

Concluding this comparative evaluation, we consider CougaR
to be the winner as it combines the fastest relaying of blocks with
the highest security against eclipse attacks, outperforming
other protocols' trade-off between speed and security in both
these dimensions.

\section{Conclusions}
\label{sec:conclusions}

We presented CougaR:
a simple but efficient, eclipse-resistant, decentralized protocol that decides
which neighbors a node should connect to in order to reduce the block
dissemination time in blockchain networks.
The two main ingredients of CougaR's link selection policy are \emph{proximity},
in terms of network latency, and \emph{randomness}, which is also crucial for
maintaining the entire overlay network in a single, connected, robust, and
low-diameter component.
To the best of our knowledge, CougaR constitutes the best solution in the
trade-off between fast and secure (eclipse-resistant) dissemination of blocks.

Along these lines we highlighted the importance of combining close and random
links, and we explored the ratio in which they should be mixed.
We also investigated the extent to which pushing node degrees higher improves
dissemination, and we concluded that it is not the number of links that warrant
a fast and reliable dissemination, but rather their educated selection.
Subsequently, we investigated the trade-off between fast, reliable, and secure
dissemination of blocks, and bandwidth consumption by tuning the level of
parallelism in body pull requests.
Finally, we compared CougaR against a set of representative state-of-the-art
dissemination algorithms for blockchain networks, assuming both a uniform and an
exponential mining power distribution, both in error-free and in faulty network
settings.

\section*{Acknowledgments}

Work funded by \emph{Input Output Hong Kong} (\emph{IOHK}) in the context of the \emph{``Eclipse-Resistant Network Overlays for
Fast Data Dissemination''} project, aiming at optimizing and securing Cardano's network overlay.
We gratefully thank IOHK for their support, as well as IOHK's engineering team for fruitful discussions and valuable feedback.

\bibliographystyle{plain}
\bibliography{./refs}

\begin{thebibliography}{10}

\bibitem{falcon}
Falcon, 2021.
\newblock \url{https://www.falcon-net.org/}.

\bibitem{segWit}
Bip: 141, 2022.
\newblock \url{https://github.com/bitcoin/bips/blob/master/bip-0141.mediawiki}.

\bibitem{cmpctBlock}
Bip: 152, 2022.
\newblock \url{https://github.com/bitcoin/bips/blob/master/bip-0152.mediawiki}.

\bibitem{bitcoinCash}
Bitcoin-cash, 2022.
\newblock \url{https://bitcoincash.org/}.

\bibitem{bitnodes}
Bitnodes, 2022.
\newblock \url{https://bitnodes.io/}.

\bibitem{byteball}
Byteball, 2022.
\newblock \url{https://obyte.org/}.

\bibitem{cardano}
Cardano, 2022.
\newblock \url{https://cardano.org/}.

\bibitem{cosmos}
Cosmos, 2022.
\newblock \url{https://cosmos.network/}.

\bibitem{fibre}
Fibre, 2022.
\newblock \url{https://bitcoinfibre.org/}.

\bibitem{iota}
Iota, 2022.
\newblock \url{https://www.iota.org/}.

\bibitem{liquidityNet}
Liquidity network, 2022.
\newblock \url{https://liquidity.network/}.

\bibitem{peernet}
{Peer-Net Simulator}, 2022.
\newblock \url{https://github.com/PeerNet}.

\bibitem{raidenNet}
Raiden network, 2022.
\newblock \url{https://raiden.network/}.

\bibitem{tps}
Scalability, 2022.
\newblock \url{https://en.bitcoin.it/wiki/Scalability}.

\bibitem{tor}
Tor network, 2022.
\newblock \url{https://www.torproject.org/}.

\bibitem{peggedSidechains}
Adam Back, Matt Corallo, Luke Dashjr, Mark Friedenbach, Gregory Maxwell, Andrew
  Miller, Andrew Poelstra, Jorge Tim{\'o}n, and Pieter Wuille.
\newblock Enabling blockchain innovations with pegged sidechains.
\newblock 2014.

\bibitem{snow}
Iddo Bentov, Rafael Pass, and Elaine Shi.
\newblock Snow white: Provably secure proofs of stake.
\newblock {\em IACR Cryptol. ePrint Arch.}, 2016:919, 2016.

\bibitem{randomGraphs}
B{\'e}la Bollob{\'a}s.
\newblock Random graphs.
\newblock In {\em Modern graph theory}. Springer, 1998.

\bibitem{ethereum}
Vitalik Buterin et~al.
\newblock A next-generation smart contract and decentralized application
  platform.
\newblock {\em white paper}, 3(37), 2014.

\bibitem{hydra}
Manuel~MT Chakravarty, Sandro Coretti, Matthias Fitzi, Peter Gazi, Philipp
  Kant, Aggelos Kiayias, and Alexander Russell.
\newblock Hydra: Fast isomorphic state channels.
\newblock {\em Cryptology ePrint Archive}, 2020.

\bibitem{propagationInBitcoin}
Christian Decker and Roger Wattenhofer.
\newblock Information propagation in the bitcoin network.
\newblock In {\em IEEE P2P 2013 Proceedings}, pages 1--10. IEEE, 2013.

\bibitem{sybil}
John~R Douceur.
\newblock {The Sybil Attack}.
\newblock In {\em International workshop on peer-to-peer systems}, pages
  251--260. Springer, 2002.

\bibitem{bitcoinNG}
Ittay Eyal, Adem~Efe Gencer, Emin~G{\"u}n Sirer, and Robbert Van~Renesse.
\newblock Bitcoin-ng: A scalable blockchain protocol.
\newblock In {\em 13th $\{$USENIX$\}$ symposium on networked systems design and
  implementation ($\{$NSDI$\}$ 16)}, pages 45--59, 2016.

\bibitem{delayAttack}
Arthur Gervais, Hubert Ritzdorf, Ghassan~O Karame, and Srdjan Capkun.
\newblock Tampering with the delivery of blocks and transactions in bitcoin.
\newblock In {\em Proceedings of the 22nd ACM SIGSAC Conference on Computer and
  Communications Security}, pages 692--705, 2015.

\bibitem{algorand}
Yossi Gilad, Rotem Hemo, Silvio Micali, Georgios Vlachos, and Nickolai
  Zeldovich.
\newblock Algorand: Scaling byzantine agreements for cryptocurrencies.
\newblock In {\em Proceedings of the 26th Symposium on Operating Systems
  Principles}, pages 51--68, 2017.

\bibitem{eclipseBitcoin}
Ethan Heilman, Alison Kendler, Aviv Zohar, and Sharon Goldberg.
\newblock Eclipse attacks on bitcoin’s peer-to-peer network.
\newblock In {\em 24th $\{$USENIX$\}$ Security Symposium ($\{$USENIX$\}$
  Security 15)}, pages 129--144, 2015.

\bibitem{peerSampling}
M{\'a}rk Jelasity, Spyros Voulgaris, Rachid Guerraoui, Anne-Marie Kermarrec,
  and Maarten Van~Steen.
\newblock Gossip-based peer sampling.
\newblock {\em ACM Transactions on Computer Systems (TOCS)}, 25(3):8--es, 2007.

\bibitem{kermarrec}
A-M Kermarrec, Laurent Massouli{\'e}, and Ayalvadi~J. Ganesh.
\newblock Probabilistic reliable dissemination in large-scale systems.
\newblock {\em IEEE Transactions on Parallel and Distributed systems},
  14(3):248--258, 2003.

\bibitem{ouroboros}
Aggelos Kiayias, Alexander Russell, Bernardo David, and Roman Oliynykov.
\newblock Ouroboros: A provably secure proof-of-stake blockchain protocol.
\newblock In {\em Annual International Cryptology Conference}, pages 357--388.
  Springer, 2017.

\bibitem{bloxroute}
Uri Klarman, Soumya Basu, Aleksandar Kuzmanovic, and Emin~G{\"u}n Sirer.
\newblock bloxroute: A scalable trustless blockchain distribution network
  whitepaper.
\newblock {\em IEEE Internet of Things Journal}, 2018.

\bibitem{kleinberg}
Jon~M Kleinberg.
\newblock Navigation in a small world.
\newblock {\em Nature}, 406(6798):845--845, 2000.

\bibitem{byzCoin}
Eleftherios~Kokoris Kogias, Philipp Jovanovic, Nicolas Gailly, Ismail Khoffi,
  Linus Gasser, and Bryan Ford.
\newblock Enhancing bitcoin security and performance with strong consistency
  via collective signing.
\newblock In {\em 25th $\{$usenix$\}$ security symposium ($\{$usenix$\}$
  security 16)}, pages 279--296, 2016.

\bibitem{omniledger}
Eleftherios Kokoris-Kogias, Philipp Jovanovic, Linus Gasser, Nicolas Gailly,
  Ewa Syta, and Bryan Ford.
\newblock Omniledger: A secure, scale-out, decentralized ledger via sharding.
\newblock In {\em 2018 IEEE Symposium on Security and Privacy (SP)}, pages
  583--598. IEEE, 2018.

\bibitem{byzantine}
Leslie Lamport, Robert Shostak, and Marshall Pease.
\newblock The byzantine generals problem.
\newblock In {\em Concurrency: the Works of Leslie Lamport}, pages 203--226.
  2019.

\bibitem{nano}
Colin LeMahieu.
\newblock Nano: A feeless distributed cryptocurrency network.
\newblock 2018.

\bibitem{dagcoin}
Sergio~Demian Lerner.
\newblock Dagcoin: a cryptocurrency without blocks.
\newblock 2015.

\bibitem{elastico}
Loi Luu, Viswesh Narayanan, Chaodong Zheng, Kunal Baweja, Seth Gilbert, and
  Prateek Saxena.
\newblock A secure sharding protocol for open blockchains.
\newblock In {\em Proceedings of the 2016 ACM SIGSAC Conference on Computer and
  Communications Security}, pages 17--30, 2016.

\bibitem{perigee}
Yifan Mao, Soubhik Deb, Shaileshh~Bojja Venkatakrishnan, Sreeram Kannan, and
  Kannan Srinivasan.
\newblock Perigee: Efficient peer-to-peer network design for blockchains.
\newblock In {\em Proceedings of the 39th Symposium on Principles of
  Distributed Computing}, pages 428--437, 2020.

\bibitem{eclipseEthereum}
Yuval Marcus, Ethan Heilman, and Sharon Goldberg.
\newblock Low-resource eclipse attacks on ethereum's peer-to-peer network.
\newblock {\em IACR Cryptol. ePrint Arch.}, 2018.

\bibitem{kademlia}
Petar Maymounkov and David Mazieres.
\newblock Kademlia: A peer-to-peer information system based on the xor metric.
\newblock In {\em International Workshop on Peer-to-Peer Systems}, pages
  53--65. Springer, 2002.

\bibitem{bitcoinTopology}
Andrew Miller, James Litton, Andrew Pachulski, Neal Gupta, Dave Levin, Neil
  Spring, and Bobby Bhattacharjee.
\newblock Discovering bitcoin’s public topology and influential nodes.
\newblock {\em et al}, 2015.

\bibitem{peersim}
Alberto Montresor and M{\'a}rk Jelasity.
\newblock Peersim: A scalable p2p simulator.
\newblock In {\em 2009 IEEE Ninth International Conference on Peer-to-Peer
  Computing}, pages 99--100. IEEE, 2009.

\bibitem{bitcoin}
Satoshi Nakamoto.
\newblock Bitcoin: A peer-to-peer electronic cash system.
\newblock Technical report, Manubot, 2019.

\bibitem{wondernetwork}
{Paul Reinheimer}.
\newblock {A day in the life of the Internet (WonderProxy)}, 2020.
\newblock
  \url{https://wonderproxy.com/blog/a-day-in-the-life-of-the-internet/}.

\bibitem{plasma}
Joseph Poon and Vitalik Buterin.
\newblock Plasma: Scalable autonomous smart contracts.
\newblock {\em White paper}, pages 1--47, 2017.

\bibitem{lightningNet}
Joseph Poon and Thaddeus Dryja.
\newblock The bitcoin lightning network: Scalable off-chain instant payments,
  2016.

\bibitem{kadcast}
Elias Rohrer and Florian Tschorsch.
\newblock Kadcast: A structured approach to broadcast in blockchain networks.
\newblock In {\em Proceedings of the 1st ACM Conference on Advances in
  Financial Technologies}, pages 199--213, 2019.

\bibitem{spectre}
Yonatan Sompolinsky, Yoad Lewenberg, and Aviv Zohar.
\newblock Spectre: A fast and scalable cryptocurrency protocol.
\newblock {\em IACR Cryptol. ePrint Arch.}, 2016:1159, 2016.

\bibitem{phantom}
Yonatan Sompolinsky and Aviv Zohar.
\newblock Phantom.
\newblock {\em IACR Cryptology ePrint Archive, Report 2018/104}, 2018.

\bibitem{tendrilstaller}
Matthew Walck, Ke~Wang, and Hyong~S Kim.
\newblock Tendrilstaller: Block delay attack in bitcoin.
\newblock In {\em 2019 IEEE International Conference on Blockchain}, pages
  1--9. IEEE, 2019.

\bibitem{monoxide}
Jiaping Wang and Hao Wang.
\newblock Monoxide: Scale out blockchains with asynchronous consensus zones.
\newblock In {\em 16th $\{$USENIX$\}$ Symposium on Networked Systems Design and
  Implementation ($\{$NSDI$\}$ 19)}, pages 95--112, 2019.

\bibitem{polkadot}
Gavin Wood.
\newblock Polkadot: Vision for a heterogeneous multi-chain framework.
\newblock {\em White Paper}, 2016.

\bibitem{rapidchain}
Mahdi Zamani, Mahnush Movahedi, and Mariana Raykova.
\newblock Rapidchain: Scaling blockchain via full sharding.
\newblock In {\em Proceedings of the 2018 ACM SIGSAC Conference on Computer and
  Communications Security}, pages 931--948, 2018.

\end{thebibliography}

\end{document}